\documentclass[journal=ancac3,manuscript=article]{achemso}
\setkeys{acs}{articletitle=true,etalmode=truncate,maxauthors=20}

\usepackage[version=3]{mhchem} 
\usepackage[T1]{fontenc}       
\usepackage{graphicx}
\usepackage{color}
\usepackage{float}

\usepackage{pdfpages}

\author{Geoffroy Kremer}
\affiliation[IJL]
{Institut Jean Lamour, UMR 7198, CNRS-Universit\'{e} de Lorraine, Campus ARTEM, 2 all\'{e}e
Andr\'{e} Guinier, BP 50840, 54011 Nancy, France}
\email{geoffroy.kremer@univ-lorraine.fr}

\author{Juan Camilo Alvarez-Quiceno}
\affiliation[CEA]
{Laboratoire de simulation atomistique, Univ. Grenoble Alpes \& CEA 38054 Grenoble France}

\author{Simone Lisi}
\affiliation[NEEL]
{Univ. Grenoble Alpes, CNRS, Grenoble INP, Institut N\'{e}el, 38000 Grenoble, France}

\author{Thomas Pierron}
\affiliation[IJL]
{Institut Jean Lamour, UMR 7198, CNRS-Universit\'{e} de Lorraine, Campus ARTEM, 2 all\'{e}e
Andr\'{e} Guinier, BP 50840, 54011 Nancy, France}

\author{C\'{e}sar Gonz\'{a}lez Pascual}
\affiliation[MADRID]{Departamento de F\'{i}sica Te\'{o}rica de la Materia Condensada and Condensed Matter Physics Center (IFIMAC), Facultad de Ciencias, Universidad Aut\'{o}noma de Madrid, E-28049 Madrid, Spain}

\author{Muriel Sicot}
\affiliation[IJL]
{Institut Jean Lamour, UMR 7198, CNRS-Universit\'{e} de Lorraine, Campus ARTEM, 2 all\'{e}e
Andr\'{e} Guinier, BP 50840, 54011 Nancy, France}

\author{Bertrand Kierren}
\affiliation[IJL]
{Institut Jean Lamour, UMR 7198, CNRS-Universit\'{e} de Lorraine, Campus ARTEM, 2 all\'{e}e
Andr\'{e} Guinier, BP 50840, 54011 Nancy, France}

\author{Daniel Malterre}
\affiliation[IJL]
{Institut Jean Lamour, UMR 7198, CNRS-Universit\'{e} de Lorraine, Campus ARTEM, 2 all\'{e}e
Andr\'{e} Guinier, BP 50840, 54011 Nancy, France}

\author{Julien Rault}
\affiliation[SOLEIL]
{Synchrotron SOLEIL, Saint-Aubin, BP 48, F-91192 Gif-sur-Yvette Cedex, France}

\author{Patrick Le F\`evre}
\affiliation[SOLEIL]
{Synchrotron SOLEIL, Saint-Aubin, BP 48, F-91192 Gif-sur-Yvette Cedex, France}

\author{Fran\c{c}ois Bertran}
\affiliation[SOLEIL]
{Synchrotron SOLEIL, Saint-Aubin, BP 48, F-91192 Gif-sur-Yvette Cedex, France}

\author{Yannick J. Dappe}
\affiliation[CEA SPEC]
{SPEC, CEA, CNRS, Universit\'{e} Paris-Saclay, CEA Saclay, 91191 Gif-sur-Yvette Cedex, France}

\author{Johann Coraux}
\affiliation[NEEL]
{Univ. Grenoble Alpes, CNRS, Grenoble INP, Institut N\'{e}el, 38000 Grenoble, France}

\author{Pascal Pochet}
\affiliation[CEA]
{Laboratoire de simulation atomistique, Univ. Grenoble Alpes \& CEA 38054 Grenoble France}

\author{Yannick Fagot-Revurat}
\affiliation[IJL]
{Institut Jean Lamour, UMR 7198, CNRS-Universit\'{e} de Lorraine, Campus ARTEM, 2 all\'{e}e
Andr\'{e} Guinier, BP 50840, 54011 Nancy, France}

\title[Defect-driven properties]
  {Electronic band structure of ultimately thin silicon oxide on Ru(0001)}


\begin{document}


\begin{abstract}
Silicon oxide can be formed in a crystalline form, when prepared on a metallic substrate. It is a candidate support catalyst and possibly the ultimately-thin version of a dielectric host material for two-dimensional materials (2D) and heterostructures. We determine the atomic structure and chemical bonding of  the ultimately thin version of the oxide, epitaxially grown on Ru(0001). In particular, we establish the existence of two sub-lattices defined by metal-oxygen-silicon bridges involving inequivalent substrate sites. We further discover four electronic bands below Fermi level, at high binding energies, two of them forming a Dirac cone at K point, and two others forming semi-flat bands. While the latter two correspond to hybridized states between the oxide and the metal, the former relate to the topmost silicon-oxygen plane, which is not directly coupled to the substrate. Our analysis is based on high resolution X-ray photoelectron spectroscopy, angle-resolved photoemission spectroscopy, scanning tunneling microscopy, and density functional theory calculations.
\end{abstract}


\section*{Introduction}

Ultrathin oxide films are of paramount technological importance in diverse fields such as catalysis or energy conversion \cite{pacchioni_cluster_1992,granqvist_transparent_2007,freund_oxide_2012,kumar_photochemical_2012}. Moreover, the continued trend towards miniaturization of modern micro- and nano-electronics has been driven significant effort in the elaboration of very-high-quality crystalline oxide films down to the ultimate thickness of a single atom or polyhedron. At this  (2D)  limit, chemical and physical properties such as energy band gap \cite{schintke}, reactivity or flexibility can be profoundly altered. As interestingly as the oxide itself, the interface with the support can exhibit unprecedented electronic properties. \\

Recently, metal-supported crystalline silicon oxide films have been grown as thin as mono- and bilayers \cite{kundu_growth_2002,schroeder_morphological_2002,weissenrieder_atomic_2005,wendt_electronic_2005,sierka_interplay_2006,todorova_atomic_2006,kaya_geometrical_2007,zhang_preparation_2008,loffler_growth_2010,lichtenstein_atomic_2012,lichtenstein_probing_2012,wlodarczyk_tuning_2012,yang_thin_2012,yu_support_2012,altman_growth_2013,shaikhutdinov_ultrathin_2013,yang_patterned_2013,crampton_atomic_2015,fischer_ultrathin_2015,mathur_degenerate_2015,yang_ultrathin_2015,klemm_preparation_2016,mathur_growth_2016,buchner_two-dimensional_2017,jhang_growth_2017,li_transition_2017}. The silicon oxide structure can be described as a network of corner--sharing SiO$_{4}$ tetrahedra forming a honeycomb lattice. This ultimately thin silicate, also called 2D silicon oxide, has remarkable properties. To date, both in its monolayer (ML) and bilayer (BL) form, it exhibits the largest band gap accessible (about 6.5 eV) among 2D materials making it the ideal 2D insulator \cite{lichtenstein_probing_2012}. Mechanical transfer from the growth substrate to a new support has been recently achieved \cite{transfert}. Therefore, this material can be envisaged to be used as a stacking brick in Van der Waals heterostructures. Like graphene, its properties can be modified by doping\cite{wlodarczyk_atomic_2013,yaang_2015,li_transition_2017}, intercalation\cite{jerratsch_lithium_2009,martinez_tailoring_2009,ulrich_realization_2009,zhong_immobilization_2017}, and creation of defects\cite{ben_romdhane_situ_2013,bjorkman_defects_2013}.
Finally, it is also an ideal plateform to investigate the amorphous-crystalline phase transformations in 2D \cite{lichtenstein_atomic_2012,ben_romdhane_situ_2013,roy2019ring}, a rising research field fuelled by the prospect for applications based on switchable properties. \\

The structure of ML and BL 2D silicon oxide grown by epitaxy on the surface of metals has been experimentally and theoretically characterized in the case of Mo(112) \cite{schroeder_morphological_2002,weissenrieder_atomic_2005,wendt_electronic_2005,sierka_interplay_2006,todorova_atomic_2006,kaya_geometrical_2007} and Ru(0001) substrates \cite{loffler_growth_2010,lichtenstein_atomic_2012,lichtenstein_probing_2012,yang_patterned_2013,fischer_ultrathin_2015,mathur_degenerate_2015,yang_ultrathin_2015,klemm_preparation_2016,mathur_growth_2016,li_transition_2017}. Ru(0001) has been demonstrated to be a substrate of choice due to its intermediate oxygen affinity and small lattice mismatch with the oxide film \cite{shaikhutdinov_ultrathin_2013}. Recently, it was also used to grow both ML \cite{lewandowski_atomic_2018} and BL \cite{lewandowski_determination_2018} germania, the parent compound where Si atoms are replaced by Ge atoms. Surprisingly, to date, the electronic band structure of such films has not been directly addressed theoretically or experimentally although its knowledge is essential to understand the dielectric and transport properties. In this article, we focus on the band structure of ultimately thin silicon oxide film on Ru(0001) measured by angle-resolved photoemission spectroscopy (ARPES) and discuss the origin of the bands in the light of polarization dependent measurements and density functional theory (DFT) calculations. \\

Prior to this analysis, we first address the chemical nature of the interface between the oxide and the substrate, which will further allow us to rationalise the band structure measurements. For that purpose we use high resolution X-ray photoelectron spectroscopy (HR--XPS), here with a higher resolution than in previous experiments,\cite{loffler_growth_2010,wlodarczyk_tuning_2012} and are hence able to resolve different kinds of bonds involving chemically inequivalent atoms in the structure. \\

Next, we present our characterization of the structure of \textit{in situ}-grown 2D ML silicon oxide,  using scanning tunneling microscopy (STM) and low energy electron diffraction (LEED). Finally, the band structure has been determined using  ARPES and interpreted in the light of DFT. We notably resolve electronic bands forming a Dirac cone at K point, associated with states delocalized in the topmost Si--O plane, and semi-flat bands associated with the hybridization with the states of the substrate. \\

\section*{Results and discussion}

\begin{figure}[H]
\begin{center}
\includegraphics[width=160mm]{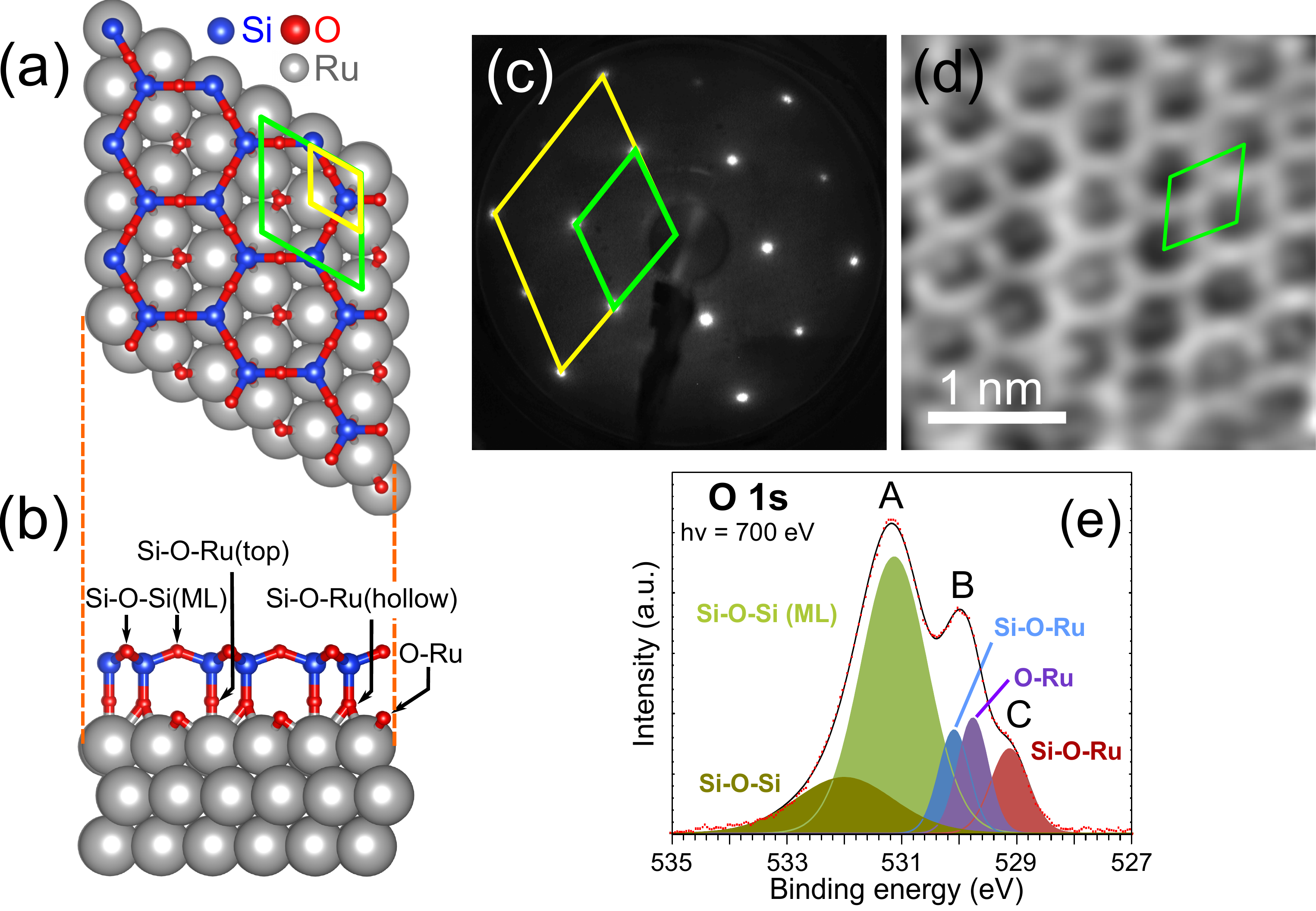} 
\end{center}
\caption{\label{Figure_structure} (a,b) Schematic structure of the ML 2D silicon oxide (top and side views, respectively) on Ru(0001).  (c) LEED pattern of ML-silicon oxide recorded at an electron energy of $80$\,eV, (d) corresponding high-resolution STM image (bias: $10$\,mV, tunneling current: $600$\,pA). (e) High-resolution XPS spectrum of the O$_{1s}$ level recorded with a photon energy of $700$\,eV. 2D silicon oxide and Ru unit cells are indicated as green and yellow rhombuses, respectively.}
\end{figure}

\textbf{Binding configuration of a monolayer 2D silicon oxide on Ru(0001).} The structural properties of the ML and BL of ultrathin silicon oxide on Ru(0001) were determined in previous works \cite{yang_thin_2012}. 2D silicon oxide is composed of SiO$_{4}$ tetrahedra forming an honeycomb-like structure whose zig-zag edges align with the [$10\bar{1}0$] direction of the surface of Ru(0001) (Figure \ref{Figure_structure}a). The lattice constant is $5.4$\,$\mathring{A}$, that is twice the one of Ru leading to a ($2\times2$) commensurate unit cell. The bonding of the ML to the support can be described by covalent Si--O--Ru bonds perpendicular to the Ru surface (Figure \ref{Figure_structure}b). In contrast, for the BL--silicon oxide no such covalent bonds exist with the substrate, and the interaction is dominated by weak Van der Waals forces. One of the evidences of the deconnexion at the BL coverage is the loss of Si--O--Ru perpendicular vibrations modes as demonstrated by Infrared Reflection Absorption Spectroscopy (IRAS) measurements and confirmed by corresponding DFT calculations.\cite{loffler_growth_2010,lichtenstein_atomic_2012,fischer_ultrathin_2015,yang_ultrathin_2015,klemm_preparation_2016,lewandowski_determination_2018} \\

In a first experimental setup, a ML-film was elaborated following the preparation method described in the Materials and Methods section. It shows a sharp ($2\times2$) LEED pattern (Figure \ref{Figure_structure}c) and a honeycomb structure in STM (Figure \ref{Figure_structure}d) with the presence of  interfacial O atoms visible as protrusions at the center of the hexagons typical of an "O-rich" phase, confirming recent results from litterature\cite{mathur_degenerate_2015,mathur_growth_2016}. This phase has been predicted as the stable phase for the ML by earlier DFT calculations \cite{shaikhutdinov_ultrathin_2013}. \\

 According to the ML structural model depicted in Figure \ref{Figure_structure}a, O atoms occupy three sites on Ru: interfacial atoms are adsorbed on Ru hollow sites (O--Ru). When bonded to Si and forming Si--O--Ru linkages, O is adsorbed on Ru hollow-- (Si-O--Ru(hollow)) or top (Si--O--Ru(top)) sites. O atoms bonded to two Si atoms are found in the topmost layer of the ML (Si--O--Si(ML)). \\

To try to detect the presence of these chemically different kinds of O atoms, we measured XPS with unprecedented high-resolution at a photon energy of $h \nu $ = 700 eV. The measured spectra exhibit three peaks (see Figure \ref{Figure_structure}e). Peak A at 531.1 eV binding energy (BE) is attributed to O bonded to two Si atoms (Si--O--Si(ML), light green) in the ML as already observed on the same system on Ru(0001)\cite{yang_thin_2012} as well as on Mo(112) \cite{kaya_geometrical_2007}. The A peak has an extended high BE-tail, which suggests that it actually comprises two components. The higher-BE component (Si--O--Si, dark green), which obviously has less spectral weight than the one around 531.1 eV, may relate most probably to bonds between Si and O atoms for instance in a small fraction of the surface being covered with a BL \cite{wlodarczyk_tuning_2012} or to locally different chemical environments corresponding to structural defects or inhomogeneities in the "free" oxygen phase on Ru(0001) coexisting with the silicon oxide. \\

In an initial fit attempt, based on XPS spectra recorded on O--p(2$\times$2), O--(2$\times$1), and 3O--(2$\times$2) superstructures on Ru(0001) (see Figure S2a), we could reasonably assign peak B located at 529.9 eV BE to O atoms chemisorbed on Ru (refered as O--Ru). Peak C is assumed to be due to Si--O--Ru bonds. Note that, up to now, peak C has not been resolved for this system\cite{yang_thin_2012}. Starting from this fitting assumption, a ratio between O--Ru and Si--O--Ru contributions of 2:1 was obtained (see Figure S3a,b for details concerning the fitting procedure) which is not consistent with the one for a Si$_4$O$_{10}$--2O formula such as expected here (with an unique Si--O--Ru site) \textit{i.e.} 1:2. Actually, to better fit the experimental data, peak B and C need to be decomposed with three contributions (see Figure S3c,d). In this case, peak B is deconvoluted into the O--Ru contribution at 529.8 eV (purple) and a second Si--O--Ru contribution at 530.1 eV (light blue) which originates from O bonded to Si and Ru. Finally, peak C is assigned to a second Si--O--Ru contribution (red). This analysis takes into account the two inequivalent sites for O in the Si--O--Ru bonds according to the structural model given in Figure \ref{Figure_structure}a. Nevertheless, we can not assign unambiguously a priori the two Si--O--Ru linkages, \textit{i.e.} whether O sits on a top or a hollow site of the substrate.  We can note that analysis from simulated XPS--spectrum reported in the Figure 10b of B. Yang \textit{et al.} \cite{yang_thin_2012} predicts that Si--O--Ru(hollow) and Si--O--Ru(top) are respectively at high and low binding energies positions. In that case the corresponding ratio extracted from XPS-spectrum for Si--O--Ru/O--Ru/Si--O--Ru contributions is equal to 1:1.1:1 which is  in good agreement with the expected one (1:1:1) for the structural model presented in Figure \ref{Figure_structure}a,b. \\

\begin{figure}[H]
\begin{center}
\includegraphics[width=160mm]{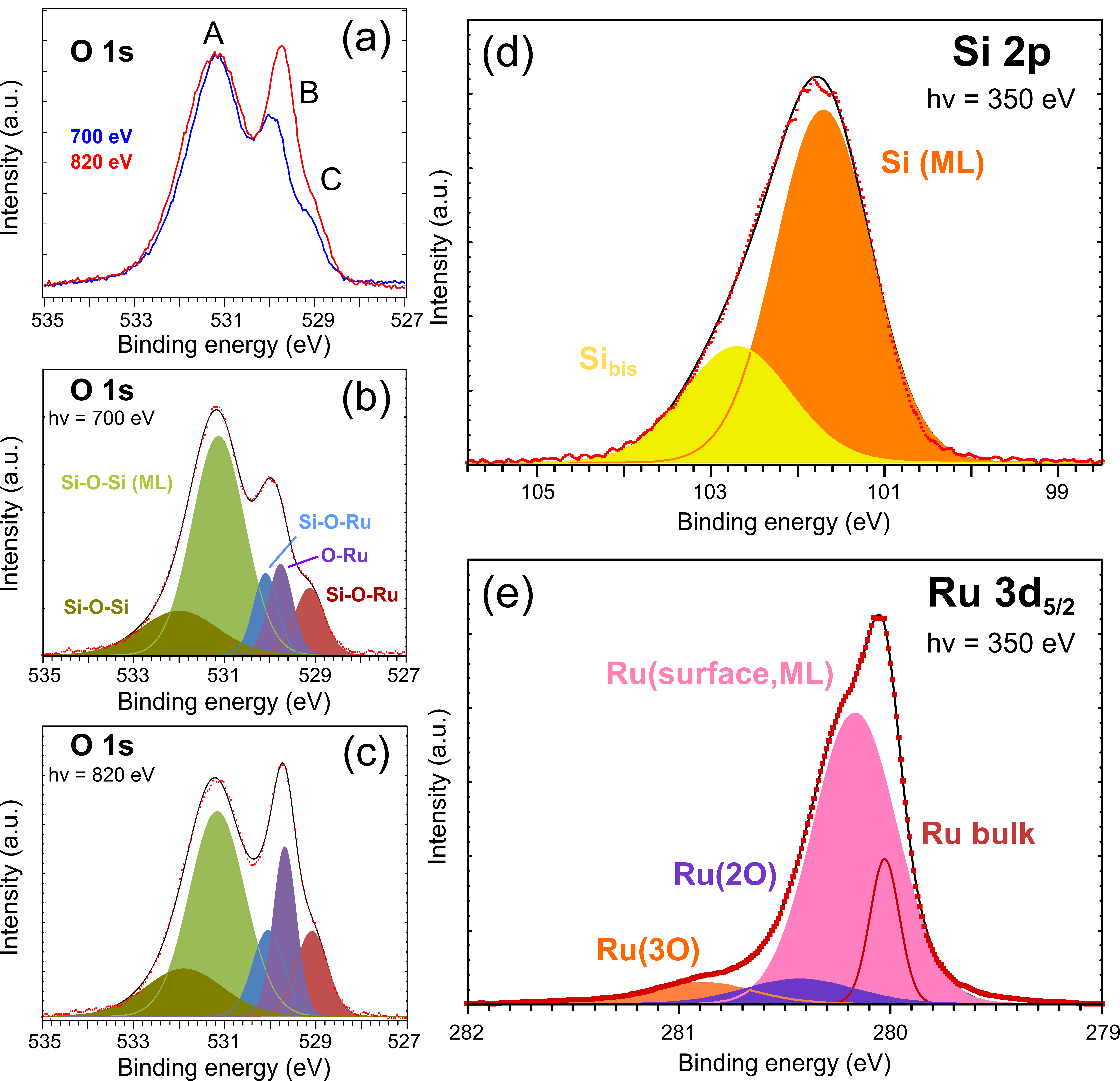}
\end{center}
\caption{\label{Figure_XPS} High-resolution XPS spectra of  ML silicon oxide over Ru(0001). (a) Raw data of O$_{1s}$ core levels recorded with a photon energy of  $700$ (blue), and $820$\,eV (red). (b,c) Corresponding fitted spectra. (d) Si$_{2p}$ and (e) Ru3d$_{5/2}$ core levels. (d) and (e) are recorded with a photon energy of $350$\,eV.}
\end{figure}

 The expected ratio between the different components corresponding to chemically inequivalent O bonds in the system (Si--O--Si(ML), Si--O--Ru, O--Ru, Si--O--Ru) is 3:1:1:1.  Our best fit to the experimental data yields a roughly twice stronger contribution for the Si--O--Si bonds (Figure S3d). The deviation from the expected ratio is due to stronger surface character of Si--O--Si bonds compared to the other ones as already observed for ML on Mo(112)\cite{kaya_geometrical_2007}  and BL on Ru(0001)\cite{wlodarczyk_tuning_2012} by varying the angle of emission with respect to the normal of the sample. One way to modify the weight of contributions with surface character is to adjust the photon energy. Here, we have increased this energy to 820 eV. In Figure \ref{Figure_XPS}a-c, we observe as expected that the Si--O--Si contribution is now significantly lower (by about 25\%), while the O--Ru contribution constituting the B peak increases (about 45\% with respect to Si--O--Ru components). \\

On the one hand, the decreased weight observed for Si--O--Si bonds can be explained by an increase of the inelastic mean free path (IMFP) of the electrons at high photon energy. On the other hand, the increased weight for O--Ru contribution has already been observed using higher photon energy, with regular X-ray lab source \cite{yang_thin_2012}. The IMFP being marginally different for the three components (O--Ru, Si--O--Ru(top), Si--O--Ru(hollow)), it cannot explain such variations. A reasonable explanation could be a photodiffraction effect that may only be accounted for within a complex quantitative analysis of the XPS spectrum. \\


We also performed XPS measurements on the Si$_{2p}$ and Ru3d$_{5/2}$ core levels (Figure \ref{Figure_XPS}d,e). The Si$_{2p}$ core level spectrum does not exhibit a single component. The spin-orbit coupling splitting of the core level, of about 600 meV, is too small to account for the lineshape. We hence assume that besides the main contribution centered at 101.7 eV (Si (ML)), at least one more contribution exists at 102.7 eV (Si$_{bis}$), corresponding to a chemically inequivalent kind of Si atoms. Similar to the case of O atoms addressed earlier, such Si atoms may be found in a small fraction of the surface being covered with a BL silicon oxide or at defect sites.\\



  We finally examine the Ru3d$_{5/2}$ core level spectra after the growth of the silicon oxide ML (Figure \ref{Figure_XPS}e and Figure S2b), and compare it to the spectra obtained after forming an oxygen surface reconstruction consisting of 3O--(2$\times$2) and (2$\times$1) domains (identified by STM, not shown here). The oxidised surface spectrum is composed of four contributions which are due to inequivalent Ru atoms at the surface and in the bulk \cite{lizzit_surface_2001}. Upon silicon oxide growth, bulk contribution remains and one extra surface component (Ru(surface,ML)) is observed at $280.2$\, eV (pink) BE (see Figure \ref{Figure_XPS}e) that can be attributed to surface Ru atoms involved in Si--O--Ru bonds. In the case of the pure ML--phase, one would expect only two components: the bulk one and the surface one assigned to Si--O--Ru. However, Ru(2O) and Ru(3O) components are observed as well, possibly originating from inequivalent Ru atoms under residual BL regions, as already discussed in the cases of Si$_{2p}$ and O$_{1s}$ core levels. \\

Overall, our detailled high resolution XPS analysis provides fine insights into the binding of silicon oxide on Ru(0001). Our data detects chemically inequivalent contributions of bonds involving oxygen, silicon, and ruthenium atoms, which could not be directly deciphered, so far, due to a limited energy resolution of the measured XPS data. In particular, we are able to confirm the binding scheme that has been proposed up to now, in which the ML silicon oxide forms two distinct bonds with Ru(0001), via Si--O--Ru bridges involving two kinds of Ru atoms, three ones in hollow sites on one hand, and a single one on top sites on the other hand. \\

\begin{figure}[H]
\begin{center}
\includegraphics[scale=0.06]{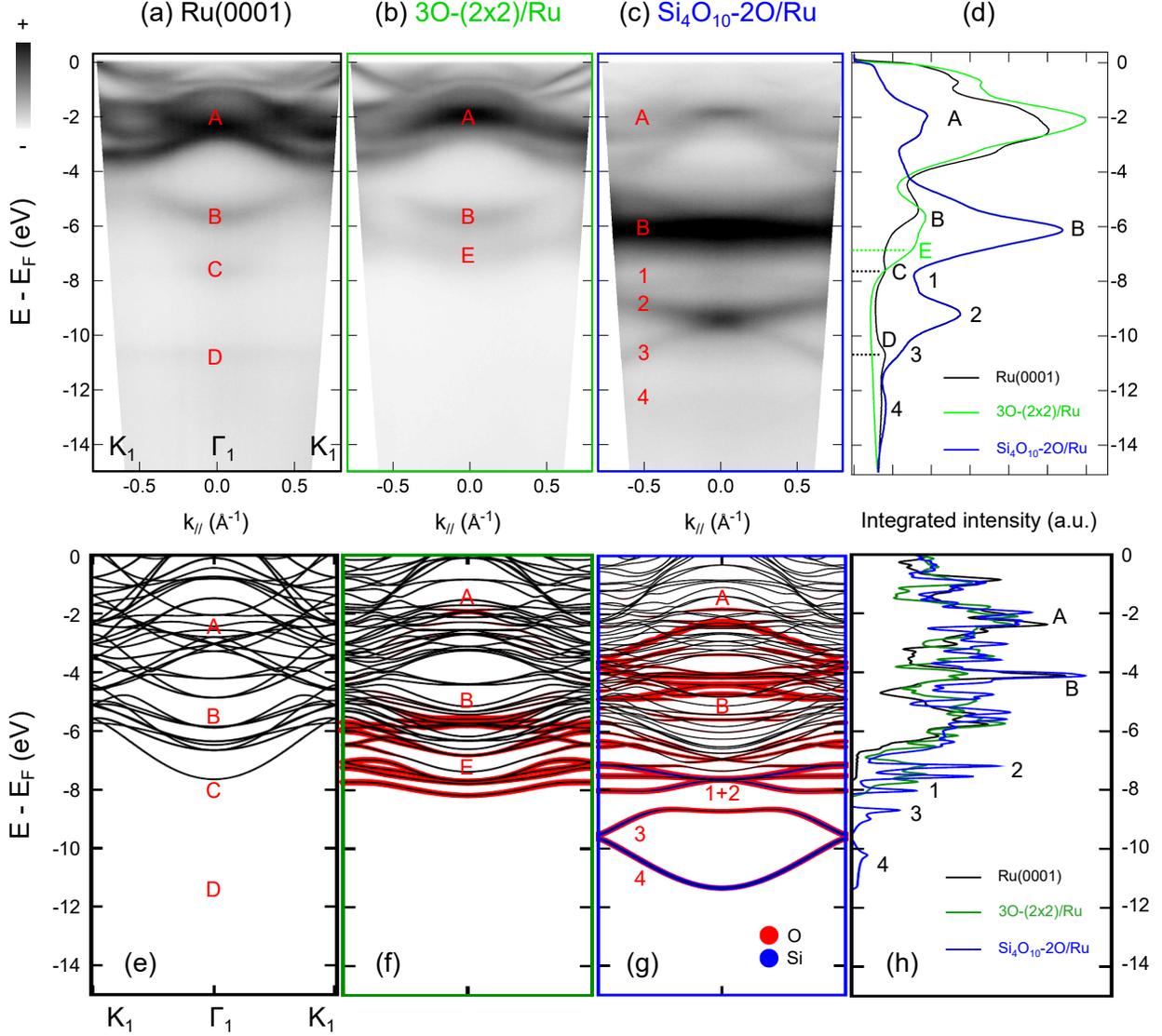} 
\end{center}
\caption{\label{Figure3} ARPES spectra using $h \nu =  40$ eV and LH polarization along the $K_{1}-\Gamma_{1}-K_{1}$ high symmetry line for (a) bare Ru(0001), (b) 3O--($2 \times 2$)/Ru(0001), and (c) ML silicon oxide. (d) Corresponding integrated intensities along k$_{\parallel}$. Corresponding DFT--calculated band structure of (e) bare Ru(0001), (f) 3O--($2 \times 2$)/Ru(0001), and (g) ML silicon oxide. Black, red and blue colors correspond respectively to Ru, O and Si character of the bands. (h) Calculated FDOS  of (e,f,g). }
\end{figure}

\textbf{Dispersive electronic states in the monolayer of ultrathin silicon oxide.} Now that we have established the chemical binding configuration at the interface between silicon oxide and Ru(0001), we turn to the exploration of the band structure of the system. Figure \ref{Figure3}a-c  displays ARPES spectra along the high symmetry $K_{1}-\Gamma_{1}-K_{1}$ line (see Figure S4 for details concerning the directions in the Brillouin zone (BZ)) of the bare substrate, the pre-oxidized 3O--(2$\times$2)/Ru(0001) and after 2D silicon oxide formation, respectively. Full density of states (FDOS) (\textit{i.e.} integrated over all the $k$ points accessible with the measurement) are given in Figure \ref{Figure3}d. The band-structure calculated with DFT for the optimized geometry corresponding to bare Ru(0001), the 3O--(2$\times$2)/Ru(0001) case, and the Si$_4$O$_{10}$--2O structural model presented in Figure \ref{Figure_structure}a,b, are displayed in Figure \ref{Figure3}d. \\

The ARPES spectrum of bare Ru(0001) (Figure \ref{Figure3}a) is characterized by (i) a group of dispersive bands "A" in the range [0,-4] eV below Fermi level (E$_{F}$), (ii) an upward dispersing band "B" at -6 eV around $\Gamma$ and (iii) a band labelled "C" at about -8 eV in good agreement with literature \cite{nguyen_electronic_2013}. These states are of sp-- and d-- like character. The flat band "D" lying at -10.5 eV  originates from residual carbon contamination; its intensity varies with surface preparation conditions. \\

Upon  O chemisorption, the ARPES spectrum is mostly modified below -6 eV where an extra band "E" is observed with a minimum at $\Gamma$ point at about -7 eV. Band "B" is marginally affected by the presence of O. In addition, we observe spectral weight near $\Gamma$ at -2 eV (band "A") and a decrease of the Ru-related intensity (compare to Figure \ref{Figure3}a) that exhibits a (2$\times$2) superperiodicity with respect to the substrate and resulting from Ru--O covalent bonds (see Figure S5). We note that the "D" band has vanished as could be expected since oxygen exposure to obtain the superstructures is  a common way to get rid of carbon contamination on Ru surfaces. Experimental data are well reproduced by the DFT calculations of both the bare and oxygen-reconstructed surfaces (Figure \ref{Figure3}e,f), specially the fact that ARPES spectrum is featureless below -8 eV. In order to simplify the interpretation, we have colorized in red the bands with an oxygen character. In comparaison with the case of the  bare surface, the calculations for the 3O--(2$\times$2) reconstruction exhibit new bands with oxygen character near -2 eV, -6 eV and -8 eV. These calculations match well with the  ARPES measurements. \\

After the growth of the silicon oxide, four bands, labelled "1" to "4" in Figure \ref{Figure3}c, are observed in the [-8,-14] eV range below E$_{F}$. Band "1" is almost flat and located at about -8 eV. Bands "2" and "3" cross at -9.5 eV and disperse downwards and upwards around  $\Gamma$ point, respectively. Band "4" disperses downwards with a minimum at $\Gamma$ at -13 eV. Note that band "4" is exclusively observed in the second BZ due to matrix elements effects as shown in  Figure S6. In the [-5,-7] eV range, additional spectral weight is now  observed. The FDOS in this range, displayed in Figure \ref{Figure3}d, is broadened. In addition to the dispersive "B" band from 3O--(2$\times$2)/Ru at -5 eV, a new flat band is found in this energy range. Moreover, a strong reduction of the photoemission intensity just below E$_{F}$ is visible (bands "A"), presumably due to the presence of the oxide limiting the escape of photoelectrons created underneath it, as it is expected given the surface sensitivity of the technique. \\ 

These features are globally reproduced by our DFT calculations displayed in Figure \ref{Figure3}g. In particular, calculations show the emergence of non dispersive states around -5 eV (bands "B") and dispersive states in the [-7,-12] eV range that are not present in the case of the bare and oxidized surfaces. These new states have an oxygen and a silicon nature (or mixed oxygen--silicon) which put in evidence by red and blue colors respectively. The interpretation of the different bands will be discussed below on the basis of light-polarization-dependent ARPES data and projected density of states (PDOS) calculations. We can already mention that the shape of the different bands is well reproduced by DFT . The relative energies of the bands are also nicely reproduced, but not their absolute positions, which are rigidly shifted by about -2 eV. A closer inspection reveals a few differences between the experimental data and the DFT calculations. First the relative positions of the bands "1" and "2" are not well reproduced by the calculations. Second the crossing between the bands "2" and "3" at the $\Gamma$ point is not found in the calculations. \\


\begin{figure}[H]
\begin{center}
\includegraphics[scale=0.475]{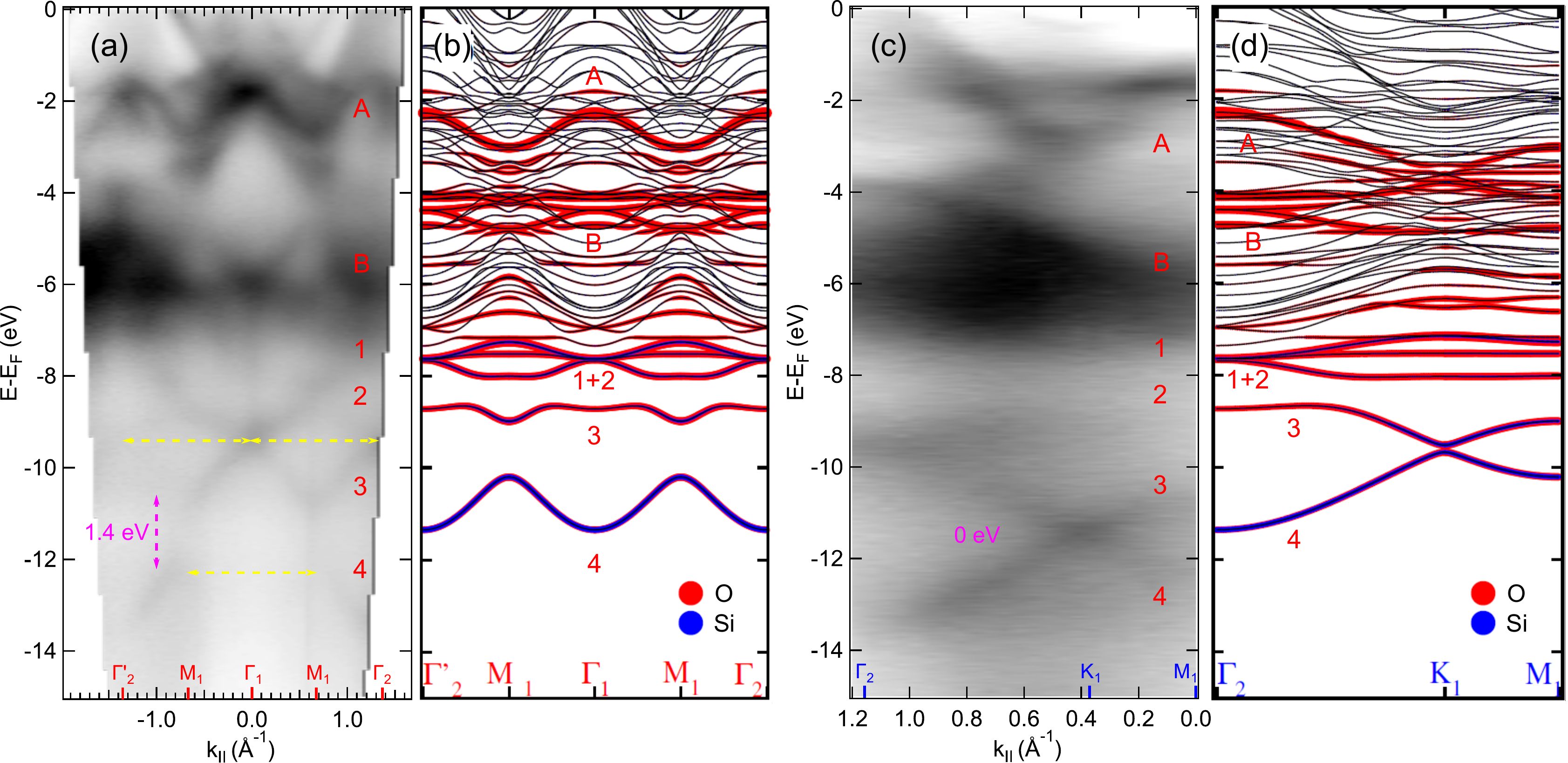} 
\end{center}
\caption{\label{Figure4} ($2 \times 2$) superperiodicity of the bands of ML silicon oxide. (a,b) Experimental and calculated bands dispersion in the $\Gamma^{'}_{2}-M_{1}-\Gamma_{1}-M_{1}-\Gamma_{2}$ high symmetry direction using LH polarization and $h \nu =  40$ eV. The distances between the high symmetry points are specified in yellow. The opening of band gap at the $M$ points is highlighted in pink. (c,d) Same for the $\Gamma_{2}-K_{1}-M_{1}$ direction using in this case LV polarization. }
\end{figure}

We also measured the band structure of silicon oxide along the  $\Gamma^{'}_{2}-M_{1}-\Gamma_{1}-M_{1}-\Gamma_{2}$ and $\Gamma_{2}-K_{1}-M_{1}$ directions (Figure \ref{Figure4}a,c). In these directions, the "A" bands show a twice smaller periodicity (here in reciprocal space) than that corresponding to the Ru(0001) lattice. These bands have the same properties in the case of Ru(0001) covered with the silicon oxide and with the 3O--(2$\times$2) (Figure S5) and are thus ascribed to the formation of O--Ru bonds. This signature in the band structure corresponds to the reminiscence of O atoms only connected to the Ru substrate in the ML structure (see Figure \ref{Figure_structure}a,b), as discussed above when analysing the HR-XPS data.  \\

Furthermore, we can again identify bands numbered "1" to "4" and the group of bands labelled "B". These bands disperse in both high symmetry directions. We mapped an extended region of reciprocal space, giving us access to high symmetry points in the second BZ of the system. We now address the superperiodicity of the silicon oxide states relative to the Ru(0001) unit cell. For both the "2-3" and "4" bands, we measure a reciprocal space periodicity of (1.34 $\pm$ 0.05) \AA$^{-1}$ (yellow dotted arrows in Figure \ref{Figure4}a), which matches the Ru(0001) lattice vector in reciprocal space ($||\Gamma_1\Gamma_2||$=$\sqrt{3}\times$0.775 \AA$^{-1}$). Concerning the Ru states, their periodicity is two times bigger and equal to (2.71 $\pm$ 0.05) \AA$^{-1}$. This point is fully coherent with the fact that silicon oxide has a ($2 \times 2$) supercell in the direct space. \\

There is overall agreement between the DFT calculations and the ARPES data acquired in the $\Gamma_1-M_1-\Gamma_2$ direction (Figure \ref{Figure4}b,d). The "3-4" bands are well reproduced. The calculated band structure of 2D silicon oxide reveals few bands with a (2$\times$2) superperiodicity that are absent in the case of pristine and oxidized Ru surfaces (see also Figure S7). Above -7 eV below E$_{F}$, a band--to--band comparison is less straightforward due to the large number of bands. \\

Interestingly, bands "3" and "4" touch at a single kind of high symmetry point in the BZ, the $K$ points (Figure~\ref{Figure4}c,d). At the vicinity of this point, the dispersion is linear (see Figure \ref{Figure4}c,d and the second derivative shown in Figure S8d), characteristic of a Dirac point. Away from these points, a non-zero bandgap exists. In fact, as can be observed in scans of the band structure along the $\Gamma-M$ and $K-M$ directions (Figure \ref{Figure4}), a saddle point is found for the "3" and "4" bands at the M points. To assess the bandgap between these two saddle points we extracted energy distribution curves (EDC) at $K$ and $M$ points (Figure S8b,c). From these curves we estimate the bandgap to (1.4 $\pm$ 0.1) eV. \\


\textbf{Origin of electronic bands.}  In the following, we address the orbital character of the electronic bands. For that purpose, we investigate the symmetry of the bands by adjusting the polarization of light (linear vertical (LV) and linear horizontal (LH)), and confront the experimental ARPES data to PDOS calculations considering the different kinds of atoms (Ru surface atoms, 4 O atoms and 2 Si atoms) composing silicon oxide as depicted in Figure \ref{Figure6}b). \\

\begin{figure}[H]
\begin{center}
\includegraphics[scale=0.15]{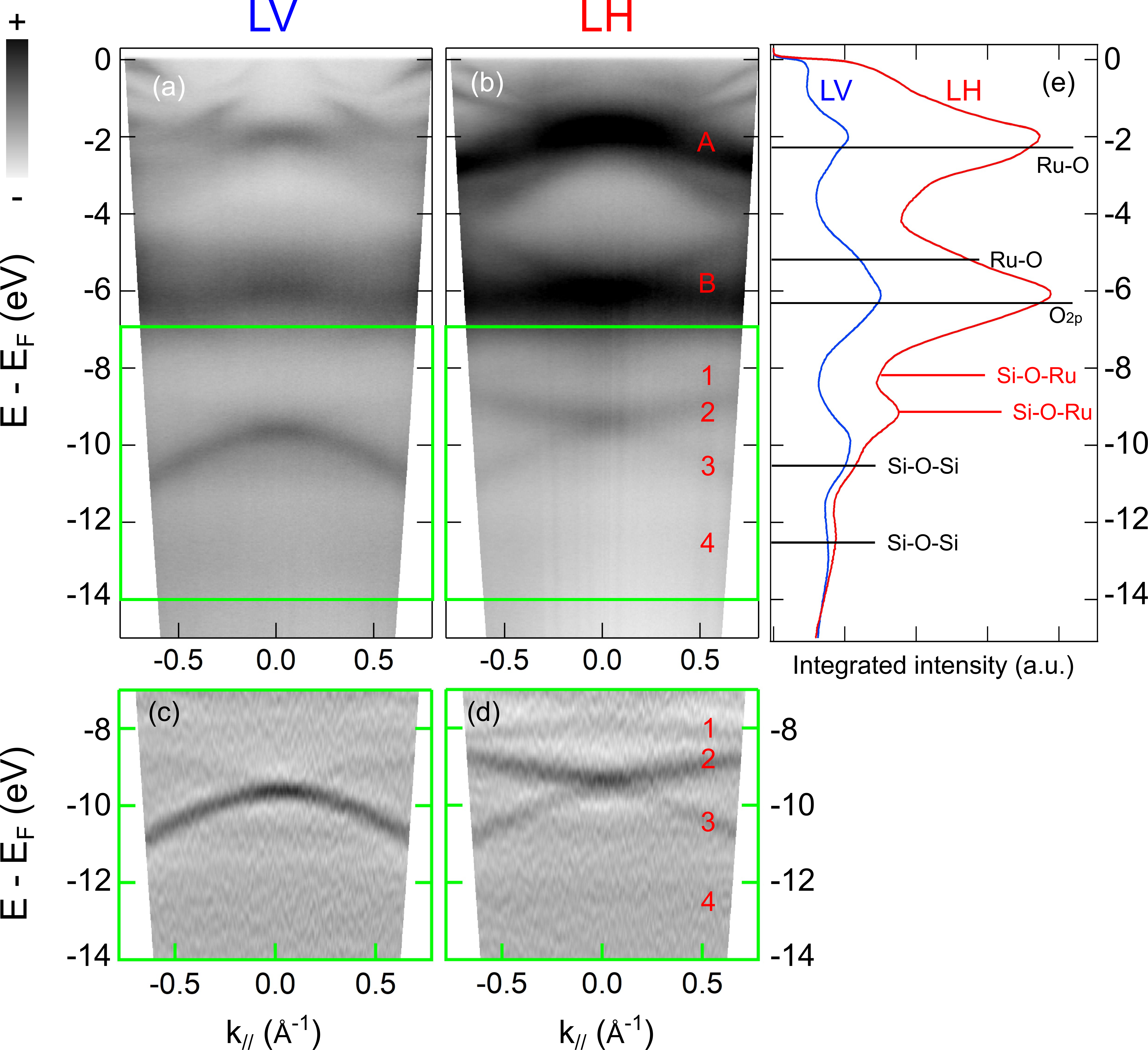} 
\end{center}
\caption{\label{Figure5} Angle-resolved spectra of ML silicon oxide in the first BZ for (a) LV and (b) LH polarization with $h \nu = 40$ eV. (c,d) corresponding second derivatives calculated in the range of [-7,-14] eV. (e) Integrated intensities of spectra shown in (a,b) along k$_{\parallel}$. Blue corresponds to LV and red to LH polarization. The origin of each band is specified by solid horizontal lines.}
\end{figure}

\begin{figure}[H]
\begin{center}
\includegraphics[width=160mm]{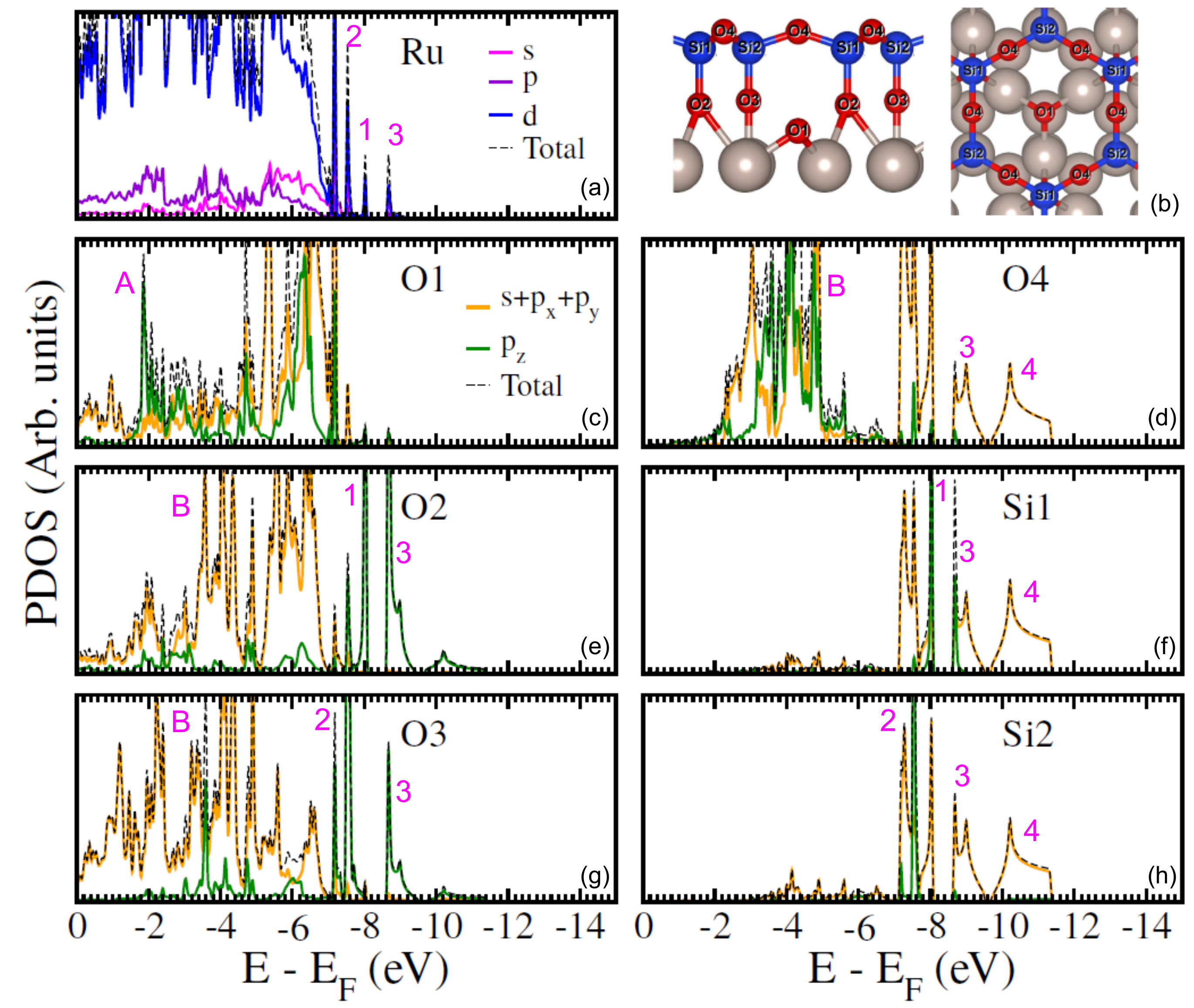} 
\end{center}
\caption{\label{Figure6} Orbital-projected density of states on O and Si atoms  in ML silicon oxide. Atoms are labeled according to the structural scheme displayed in (b).} 
\end{figure}

The ARPES spectra of ML silicon oxide, measured with both LV and LH polarizations, are shown in Figure \ref{Figure5}a-d. Second derivative is used in order to increase the visiblity of bands with a low photoemission intensity. The corresponding FDOS for both polarizations is given in Figure \ref{Figure5}e. The contributions at -2 eV ("A") and  at -5 eV (shoulder of block "B" at low BE) below E$_{F}$  are assigned to hybridised states between Ru and O \cite{wlodarczyk_tuning_2012}. Indeed these states were already present in the case of the 3O--(2$\times$2)/Ru(0001) reconstruction (see Figure \ref{Figure3}b and Figure S5). Furthermore, they mostly correspond to the density of states observed for the p$_{z}$ orbital of O$_{1}$ atom and d orbitals of Ru atoms according to our calculations in Figure \ref{Figure6}. \\

An increase of the spectral weight associated to the group of bands labeled "B" is observed in [-4,-6] eV range, compared to the case of 3O--(2$\times$2)/Ru(0001). This increase has already been observed in amorphous SiO$_{2}$ and GeO$_{2}$ \cite{distefano_photoemission_1971,fischer_electronic_1977}, in crystalline ML silicon oxide on  Mo(112) \cite{schroeder_morphological_2002,wendt_electronic_2005,kaya_geometrical_2007}, and in BL silicon oxide on  Ru(0001) \cite{wlodarczyk_tuning_2012},  and can be attributed to O$_{2p}$ non bonding states involving O atoms in Si--O--Si bonds of the topmost atomic layer. This is consistent with the PDOS of O$_{4}$ atoms which contribute to those bonds in Figure \ref{Figure6}d. Nevertheless, oxygen atoms O$_{2}$ and O$_{3}$ involved in Si--O--Ru(hollow) and Si--O--Ru(top) bonds also display non negligible PDOS in this energy range and might also participate in the total spectral weight, in particular with non bonding flat states with  p$_{x}$ and  p$_{y}$ symmetry.   \\

  The most probable origin of bands "1" and "2" is an hybridisation of p$_{z}$ orbitals  from O and Si atoms  in Si--O--Ru bonds connecting the silicon oxide sheet to the ruthenium substrate.  On the one hand, these bands are drastically dependent on polarization: they are only observed with LH polarized light (see second derivative data in Figure \ref{Figure5}c-d). This point can be understood by considering the expression of the photoemission intensity, in particular by considering parity arguments\cite{malterre,yukawa_electronic_2013} given rise to particular selection rules for a given light polarization and a given probed initial state. In this way, it is possible to demonstrate that LV polarization is in--plane sensitive but not out--of--plane sensitive and yields no signal in the latter case. On the contrary, LH is both in--plane and out--of--plane sensitive. Using these arguments, bands "1" and "2", which are drastically affected by the modification of the polarization, can be experimentally attributed to out--of--plane covalent Si--O--Ru bonds. On the other hand, our calculations exhibit  a large contribution of p$_{z}$ orbitals from Si$_{1}$/O$_{2}$ and Si$_{2}$/O$_{3}$  atoms in the [-7,-8] eV energy range.  This is confirmed by the PDOS contribution of Ru surface atoms (d orbitals) in the corresponding energy range (see Figure \ref{Figure6}a). Even if bands "1" and "2" are not well reproduced by our DFT calculations and hardly distinguishable in comparison to the ARPES data, we can still conclude on their Si--O--Ru origin. These bands are indeed absent for an O reconstruction on Ru(0001) (Figure \ref{Figure3}b) and for a BL silicon oxide (data not shown). \\

Concerning band "3", we observe photoemission intensity with both LH and LV polarizations (maximum in the case of LV). Following the same line of thoughts as before, we conclude that this band has an in--plane character. This is partially confirmed by our PDOS calculations, which indicate that this state emerges from  hybridisation in the Si--O--Si bonds at -9 eV from O$_{4}$ and Si$_{1,2}$ atoms. Nevertheless, Ru surface, O$_{2}$ and O$_{3}$ atoms also exhibit  non negligible spectral weight in this energy range which contradicts our interpretation in terms of polarization arguments. \\

We note that this state has already been observed in other silicon oxide compounds whose structure involves only Si--O--Si bonds: amorphous SiO$_{2}$ \cite{distefano_photoemission_1971} and silicon oxide BL \cite{wlodarczyk_tuning_2012} though its dispersion was not resolved so far. This last point is coherent with an interpretation of band "3" as a Si--O--Si contribution, rather than a Si--O--Ru contribution that cannot exist in the silicon oxide BL.  \\

The last band numbered "4", visible in the second BZ (Figure S8) and not in the first one (Figure \ref{Figure5}b,d), can also be assigned to Si--O--Si bonds. Indeed this band appears in both LH and LV cases. In addition, calculations predict DOS from orbitals of O$_{4}$ atoms and from orbitals of Si$_{1,2}$ atoms in the energy range of [-10,-11.5] eV below E$_{F}$. This interpretation is also corroborated by the absence of a PDOS contribution of Ru surface atoms as described in Figure \ref{Figure6}a. This band is unambiguously due to an hybridisation of O$_{4}$ and Si$_{1,2}$ orbitals at the top of the silicon oxide layer. \\

Overall, the ML silicon oxide on Ru(0001) is characterized by at least four inequivalent dispersive bands. Two of them are presumably due to out--of--plane Si--O--Ru covalent bonds connecting the silicon oxide sheet to its substrate, and two others are due to in--plane Si--O--Si bonds in the topmost plane of silicon oxide sheet. While Si--O--Si bands are well reproduced by DFT calculations using the generally accepted relaxed model shown in Figure \ref{Figure_structure}a,b, it is not the case of Si--O--Ru ones. Indeed, in addition to a global energy shift, contributions "1--2" are found at too low BE, and the degeneracy of bands "2" and "3" is lifted at the $\Gamma$ point, unlike in the ARPES measurements. \\

In the DFT calculation we have considered the atomic model initially proposed in Ref. \cite{yang_thin_2012} for the monolayer silicon oxide. The precise epitaxial relationship of this model (hollow-top) was further supported by combining DFT calculations and atomically-resolved STM images \cite{mathur_degenerate_2015} and is confirmed by the present XPS measurements (cf Figure \ref{Figure_XPS}e). Indeed,  we report two well defined contributions for the Ru--O--Si bonds which energies are separated by 1 eV. This highlights a complex bond character in the two Ru--O--Si bridges linking  the monolayer silicon oxide to the ruthenium surface.
 
One striking feature in the calculated band structure is the presence of five bands (Figure \ref{Figure3}g) while only four bands are observed in the ARPES. Moreover, band "3" is flattened in $\Gamma$ leading to an apparent gap opening due to an anti-crossing with the next bands. We can assume that for the same reason, some degeneracy are raised between two bands in the calculation leading to five bands instead of four. In other words, it is likely that band "1" or band "2" are degenerated in the experimental band structure.

In order to evaluate the chemistry of the Ru--O--Si bridges, we have tested different possible origins for the discrepancy between theoretical and experimental band structures: \textit{i/}  epitaxial relationship \textit{ii/} number of ruthenium layers that could change the electrostatics at the Ru surface.  However, all the high symmetry shifts relaxed to the present model. Likewise increasing the number of Ru layers does not allow to recover the experimental behavior at the $\Gamma$ point (See Figure S10b).
 
 An other possible source of discrepancy might come from the used level of approximation in the DFT calculations. Indeed the present system is an hybrid one that contains different type of bondings ranging from covalent to metallic and might be highly sensitive to the used approximations. This hypothesis was tested by considering a bunch of different exchange and correlation (XC) functional including local (LDA) semi-local (PBE, SCAN) and hybrid (HSE06) descriptions. In Figure S10c we report the band structure for the SCAN functional that displays a small but rigid downward shift of the five characteristic bands. All the other considered XC functionals show the same behavior as depicted in Figure S11 by the evolution of the five characteristic eigenvalues at the $\Gamma$ point. We note that the HSE06 functional seems to better reproduce the position of the eigenvalues "4" and "3" in connection with the expected band gap underestimation for local XC functionals.   
 
 At that point we have to conclude that the atomic model for monolayer silicon oxide\cite{yang_thin_2012, mathur_degenerate_2015}, misses a key feature at the level of the bridge that is present in the experimental grown sample. The resolution of this atomic model is above the scope of the present paper.

\section*{Conclusions}

In conclusion, we were able to resolve the binding configuration of ML of silicon oxide on Ru(0001). Two kinds of Si--O--Ru bridges involving two chemically inequivalent Ru atoms are formed. They result in the existence of two sublattices in the honeycomb lattice of ML silicon oxide. We discovered four electronic bands below Fermi level. Two of them form a Dirac cone, and two others are semi-flat bands. The existence of the two sub-lattices translates in the formation of a large band gap. All together the band structure is reminiscent of that typical of Kagome lattices. In addition, we demonstrate that the acknowledged atomic model for ML silicon oxide is incomplete as it presents few inconsistencies with our experimental data. Further characterizations are required to probe a better model.

Our work opens the way to the exploration of topologically non trivial electronic band structures in ultimately thin oxides which may eventually be controlled efficiently via local electric fields owing to the two-dimensional character of the system, for instance with the help of adsorbed species or dielectric gates.

\section*{Materials and methods}

\textbf{Experiment.} Experiments were carried out in three ultra--high vacuum (UHV) setups ($P < 1 \times 10^{-10}$\,mbar). The first one is equipped  with a low temperature STM (LT--STM) operating  at T$= 77$\,K, LEED, ARPES and XPS. The second one is equipped with a monochromated X-rays source (Al K$_{\alpha}$,  resolution better than 300 meV) and a high energy, momentum and spin photoemission analyser (DA30-L from VG--SCIENTA). Finally, the third one is the end station of the CASSIOPEE beamline at synchrotron SOLEIL, equipped with LEED, Auger electron spectroscopy (AES), high energy and momentum resolution ARPES (10 meV and 0.01 $\mathring{A}^{-1}$, respectively) and XPS (70 meV) using a VG--SCIENTA R$4000$ analyzer. Photoemission measurements were recorded at $300$\,K.  ARPES was performed with light polarization either linear vertical or linear horizontal.  Details about the experimental geometry are given in Figure S1 in  supporting information (SI). A clean Ru(0001) surface was obtained  by repeated cycles of Ar$^{+}$ sputtering and annealing up to $1400$\,K followed by molecular oxygen exposure and flash annealing  
resulting in a sharp ($1\times1 $) LEED pattern (not shown here). The absence of contamination was checked by XPS and ARPES.  After cleaning, Ru3d$_{5/2}$ core levels  exhibit a surface-related contribution at a binding energy of $279.8$\,eV (not shown here). A ML silicon oxide was grown on an oxygen-covered Ru(0001) surface forming a so-called 3O--(2$\times$2)\cite{kim_structural_1998} reconstruction observed with LEED, XPS, APRES and LT-STM. The latter reconstruction was obtained by exposing Ru(0001)  at $1 \times 10^{-6}$\,mbar O$_{2}$ at $625$\,K  for $10$\,min. Then silicon was evaporated  using electron bombardment of a high purity Si rod (> 99.9999 $\%$)  under an oxygen pressure of $3 \times 10^{-7}$\,mbar at room temperature (RT). The Si deposition rate was calibrated using AES for a well-documented system, Si on Cu(100) \cite{lalmi_epitaxial_2010}. The final crystallization step was performed under $3 \times 10^{-6}$\,mbar O$_{2}$ at $1125$\,K for $15$\,min followed by a slow temperature ramp at a rate of $10^{\circ}$C $\cdot$ min$^{-1}$ down to RT. Temperatures were measured using a pyrometer.

\noindent \textbf{Computational details.} 
The theoretical study was carried out by using first principles calculations based on density functional theory (DFT). 
The exchange correlation potential was treated within the Local Spin-Density L(S)DA approximation~\cite{Ceperley1980-LDA}. 
The Projected Augmented Wave (PAW) method~\cite{Blochl1994} was used to solve the Kohn-Sham equations as implemented in the Vienna {\it ab-initio} Simulation Package (VASP)~\cite{Kresse1996} with a kinetic energy cut-off for the plane-wave expansion of 490 eV. 
The Ruthenium (0001) surface cleaved from a hcp lattice where the crystal lattice parameter was firstly optimized.  We have considered three atomic layers that is expected to be enough to reproduce bulk properties of the growth support \cite{yang_thin_2012}. 
The equilibrium geometry of Oxygen adsorbed and silicon oxide supported on the Ru(0001) surface was obtained by relaxing the Oxygen/silicon oxide and external Ruthenium ion positions while the deepest Ruthenium layer was kept fixed. 
The optimization of atomic positions was performed using conjugate gradient algorithm until the Hellmann-Feynman forces reach the threshold of 0.1$\times10^{-3}$~eV/{\AA}.
A vacuum of at least 12 {\AA} was employed along the z-direction to avoid undesired interactions between periodic layers.
The Brillouin zone was sampled with a $20\times20\times1$ k-point mesh in the self-consistent energy calculations and increased to a denser $40\times40\times1$ mesh in the density of states calculation.

\begin{suppinfo}
Supporting Information includes details concerning the experimental geometry used for photoemission measurements, discussion of line shapes used in XPS data fitting procedure, surface Brillouin zone details, ARPES data for the 3O--(2$\times$2)/Ru(0001) superstructure, additional ARPES data in the second Brillouin zone for ultrathin silicon oxide and complementary band structure DFT calculations.

\end{suppinfo}

\section{Author Information}
*E-mail: geoffroy.kremer@univ-lorraine.fr

\section{Associated Content}
The authors declare no competing financial interest.

\begin{acknowledgement}

This work was supported by the 2DTransformers project under the OH-RISQUE program of the French National Research Agency (ANR-14-OHRI-0004). We would like to thank the team Da$\mu$m for the helpful assistance during the connection and the installation of the new SR--ARPES setup on the Tube. The DFT calculations were done using French supercomputers (GENCI, \# 6194) and the Predictive Simulation Center facility that gathers in Grenoble SPINTEC, L$\_$Sim and Leti. We thanks Professor N. Mousseau for useful discussions. C. G. acknowledges financial support from the Spanish Ministry of Science, Innovation and Universities through the project MAT2017-88258-R and the "Mari\'a de Maeztu" program for units of excellence in R \& D (grant no. MDM-2014-0377).

\end{acknowledgement}

\newpage

\bibliography{biblio}



\section*{Supplementary material}

# Electronic band structure of ultimately thin silicon oxide on Ru(0001)

# Supporting information

Geoffroy Kremer,[*,†] Juan Camilo Alvarez-Quiceno,[‡] Simone Lisi,[¶] Thomas Pierron,[†] César González Pascual,[§] Muriel Sicot,[†] Bertrand Kierren,[†] Daniel Malterre,[†] Julien Rault,[∥] Patrick Le Fèvre,[∥] François Bertran,[∥] Yannick J. Dappe,[⊥] Johann Coraux,[¶] Pascal Pochet,[‡] and Yannick Fagot-Revurat[†]

[†]*Institut Jean Lamour, UMR 7198, CNRS-Université de Lorraine, Campus ARTEM, 2 allée André Guinier, BP 50840, 54011 Nancy, France*

[‡]*Laboratoire de simulation atomistique, Univ. Grenoble Alpes & CEA 38054 Grenoble France*

[¶]*Univ. Grenoble Alpes, CNRS, Grenoble INP, Institut Néel, 38000 Grenoble, France*

[§]*Departamento de Física Teórica de la Materia Condensada and Condensed Matter Physics Center (IFIMAC), Facultad de Ciencias, Universidad Autónoma de Madrid, E-28049 Madrid, Spain*

[∥]*Synchrotron SOLEIL, Saint-Aubin, BP 48, F-91192 Gif-sur-Yvette Cedex, France*

[⊥]*SPEC, CEA, CNRS, Université Paris-Saclay, CEA Saclay, 91191 Gif-sur-Yvette Cedex, France*

E-mail: geoffroy.kremer@univ-lorraine.fr

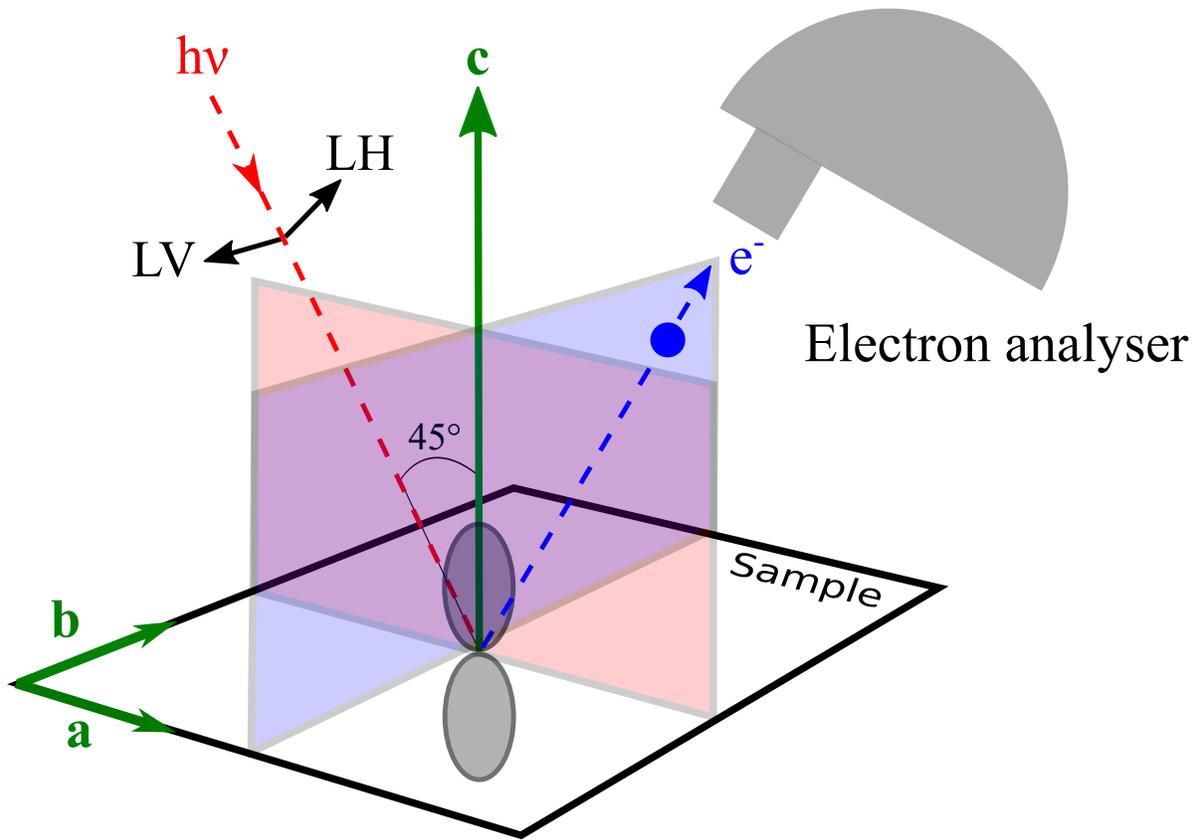

Figure S1: Experimental geometry of the photoemission experiment at the Cassiopee beamline. Red and blue planes correspond respectively to the incidence plane of photons and to the detection plane of the analyser using vertical slit. It is important to say that the selection rules mentioned in the main part of the paper are rigorously exact only at the normal emission.

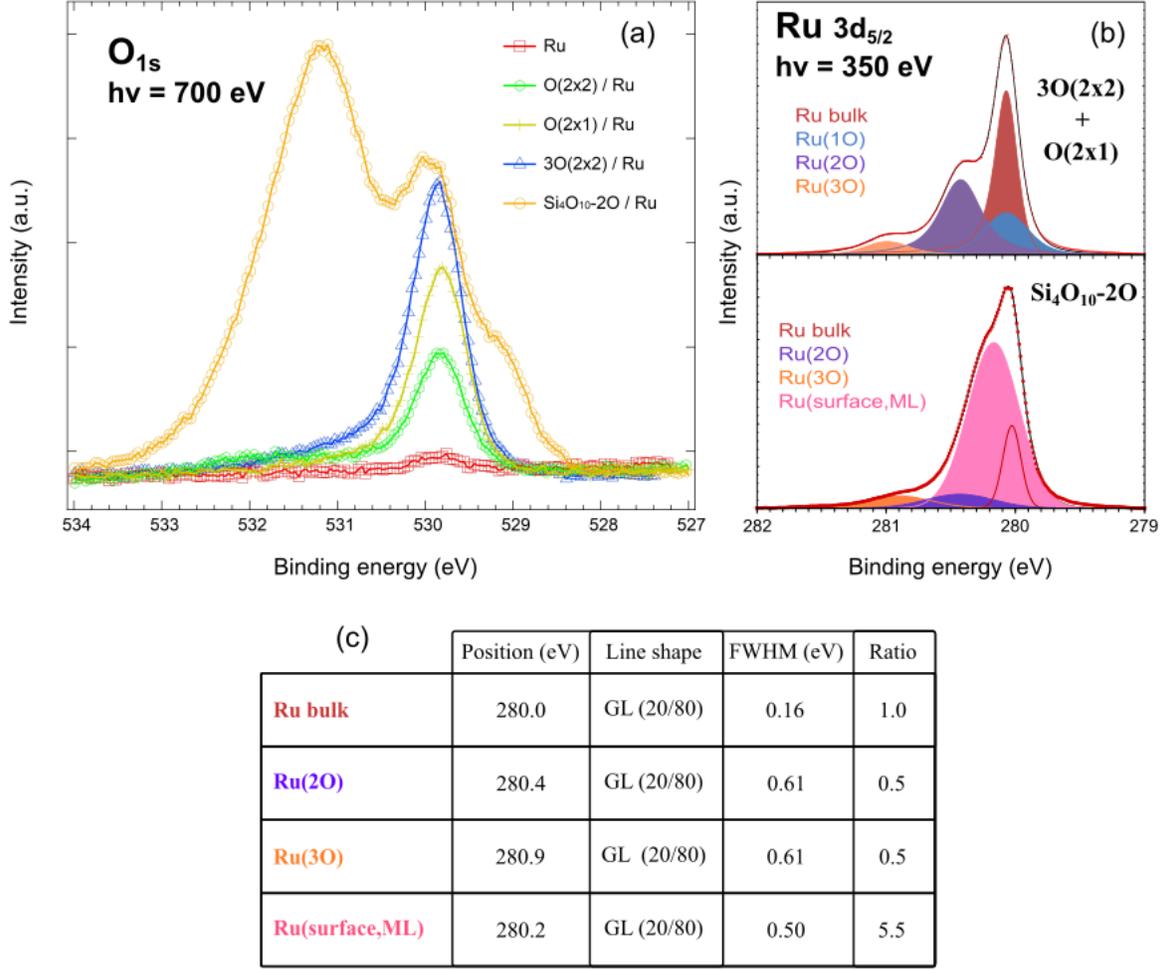

Figure S2: (a) Comparison of the $O_{1s}$ XPS core level between Ru, oxidised surfaces of Ru and ML silicon oxide. (b) Core level spectra for Ru3$d_{5/2}$, in the case of (top) a Ru(0001) surface covered with a mixture of two oxygen reconstructions (3O–(2×2) and (2×1)) and (bottom) a ML silicon oxide on Ru(0001). A fit to the data is proposed in the case of the oxygen reconstruction, for which three kinds of Ru atoms exist on the surface, with 1 (Ru(1O)), 2 (Ru(2O)) and 3 (Ru(3O)) neighbour O atoms bond to the surface in hollow sites. (c) Details concerning the fitting procedures parameters.

During the fitting procedure, a Shirley background has been used. In order to simulate the experimental core levels line shapes, we used pseudo Voigt functions GL(x/y), which correspond to a product of a Lorentzian and a Gaussian with x and y proportions respectively (an analytic form for the convolution of a Gaussian with a Lorentzian is not available). For example, GL(100) corresponds to a pure Lorentzian profile.

3

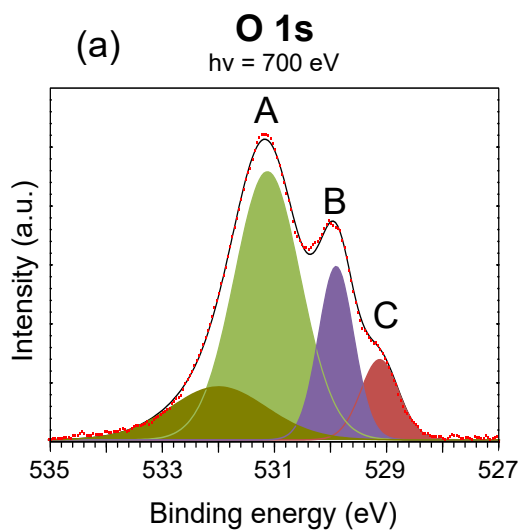
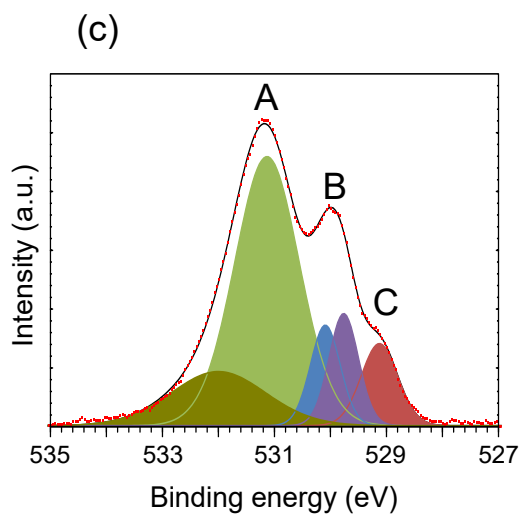

| (b) | Position (eV) | Line shape | FWHM (eV) | Ratio |
|---|---|---|---|---|
| Si-O-Si (ML) | 531.1 | GL (20/80) | 1.36 | 5.8 |
| Si-O-Ru | 530.1 | GL (20/80) | 0.63 | 0.0 |
| O-Ru | 529.9 | GL (20/80) | 0.74 | 2.1 |
| Si-O-Ru | 529.1 | GL (20/80) | 0.77 | 1.0 |
| Si-O-Si | 532.0 | GL (20/80) | 2.00 | 1.7 |

| (d) | Position (eV) | Line shape | FWHM (eV) | Ratio |
|---|---|---|---|---|
| Si-O-Si (ML) | 531.1 | GL (20/80) | 1.36 | 5.8 |
| Si-O-Ru | 530.1 | GL (20/80) | 0.63 | 1.0 |
| O-Ru | 529.8 | GL (20/80) | 0.62 | 1.1 |
| Si-O-Ru | 529.1 | GL (20/80) | 0.77 | 1.0 |
| Si-O-Si | 532.0 | GL (20/80) | 2.00 | 1.7 |

Figure S3: Fitting procedure of an high-resolution XPS spectrum of the $O_{1s}$ levels for ML silicon oxide recorded with a photon energy of 700 eV using (a) four and (c) five contributions. (b,d) Details concerning the fitting procedure.

4

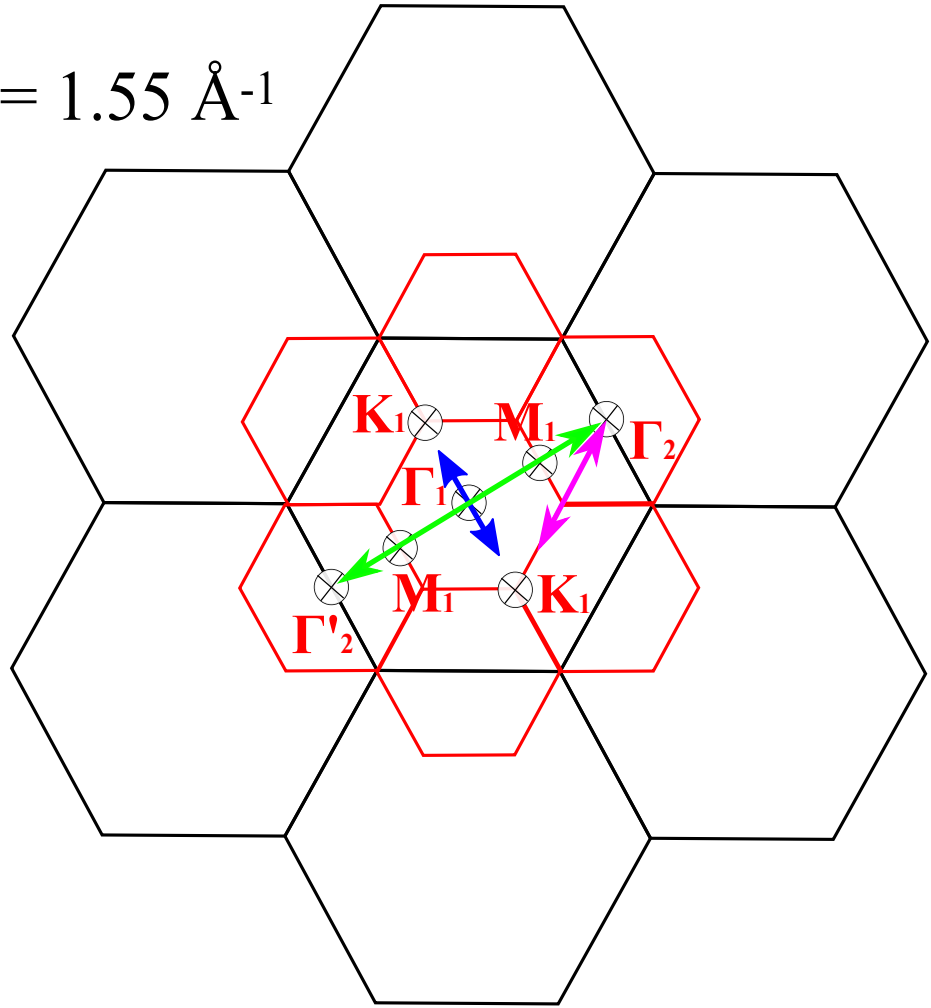

Figure S4: Representation of the extended Brillouin zones of the (1×1) of Ru(0001) (black) and (2×2) of 2D silicon oxide (red). The high symmetry points are given. Blue line correspond to scan direction in Figure 3 in the main text. Green and pink lines correspond to scan directions in Figure 4(a) and Figure 4(c) respectively in the main text.

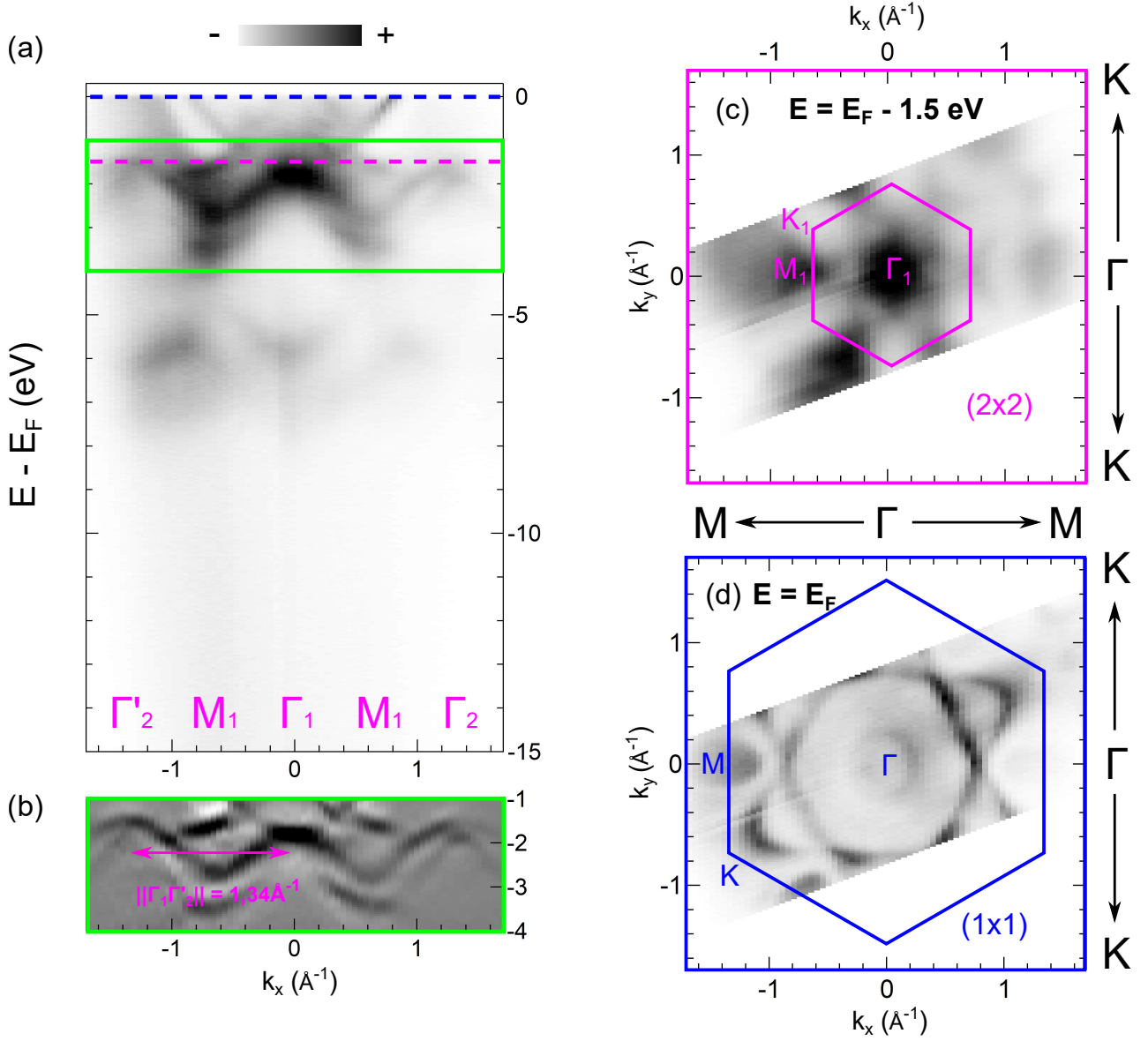

Figure S5: (a) Experimental ARPES spectrum along the $\Gamma'_2 - M_1 - \Gamma_1 - M_1 - \Gamma_2$ high symmetry direction using LH polarization and $h\nu = 40$ eV of 3O–(2×2)/Ru(0001) (b) corresponding second derivative calculated spectrum in the range of [-1,-4] eV (c,d) isoenergetic cuts at $E = E_F - 1.5$ eV and $E = E_F$ respectively. These cuts illustrate the (2×2) periodicity in the case of the Ru–O states and the (1×1) in the pure Ru states at the Fermi level.

6

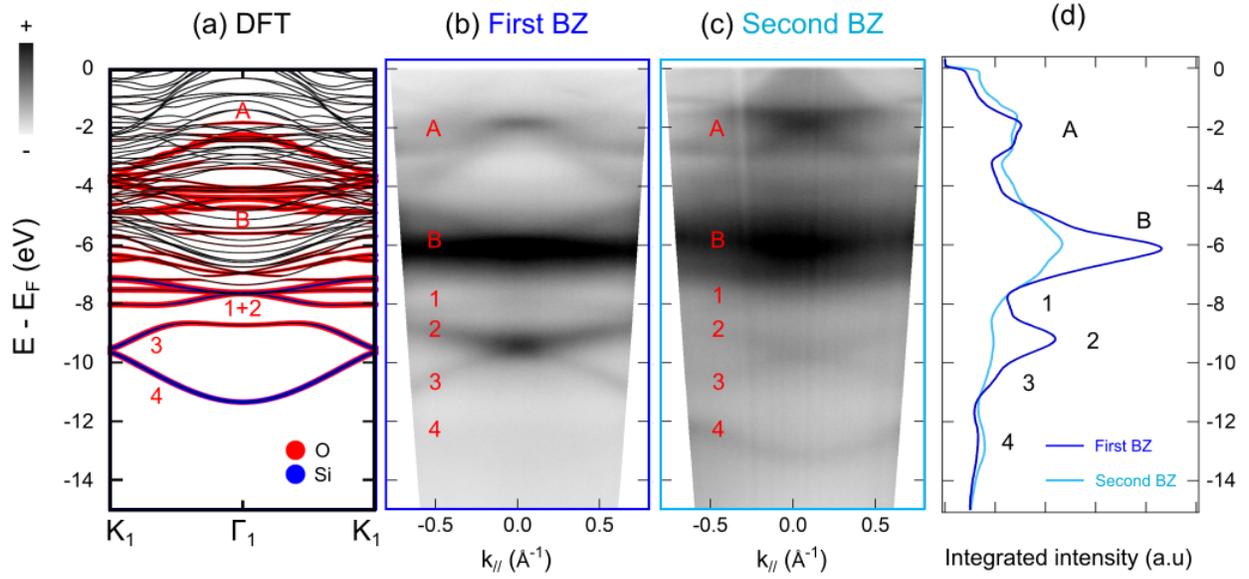

Figure S6: (a) Calculated spectrum along the $K_1 - \Gamma_1 - K_1$ high symmetry line for ML silicon oxide (b) correponding ARPES spectrum in the first BZ (c) second BZ (d) Density of states corresponding to (b,c).

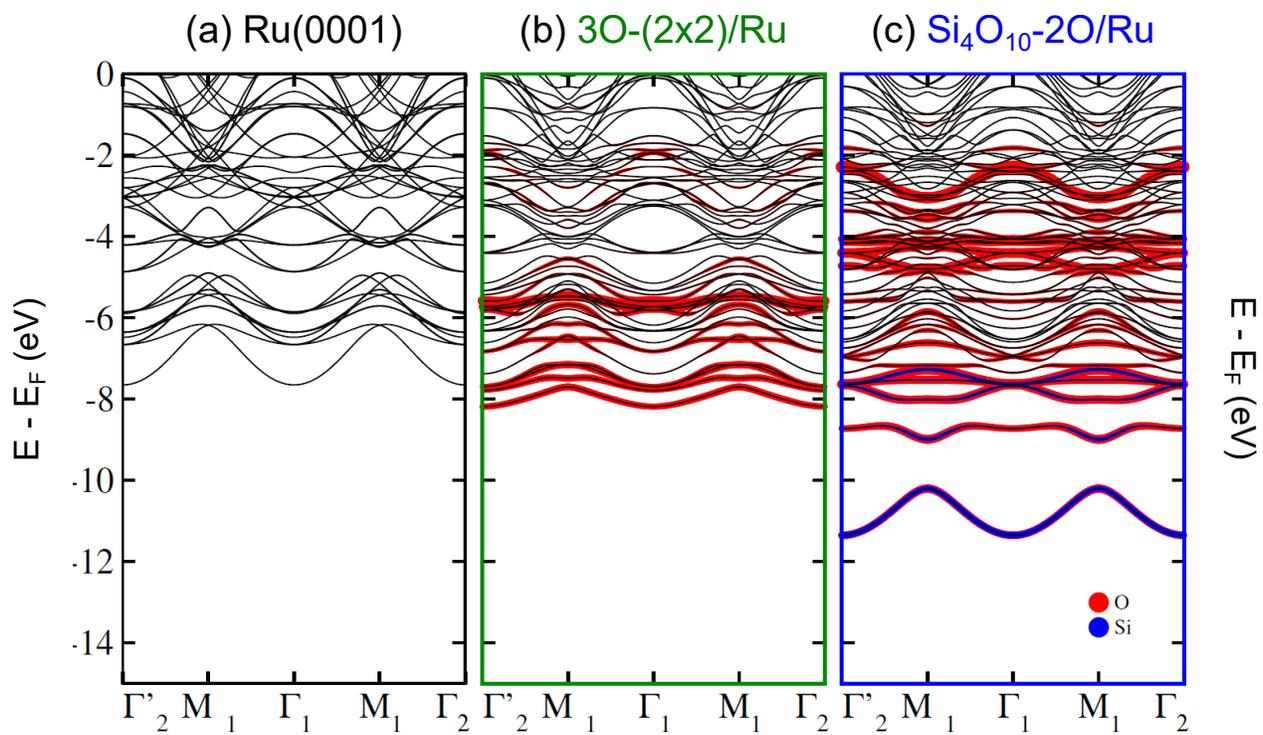

Figure S7: Calculated band structures in $\Gamma-M-\Gamma$ high symmetry line of (a) Ru(0001), (b) 3O-(2×2)-Ru(0001) and (c) ML silicon oxide on Ru(0001). Bands with oxygen and silicon character are respectively colorized in red and blue.

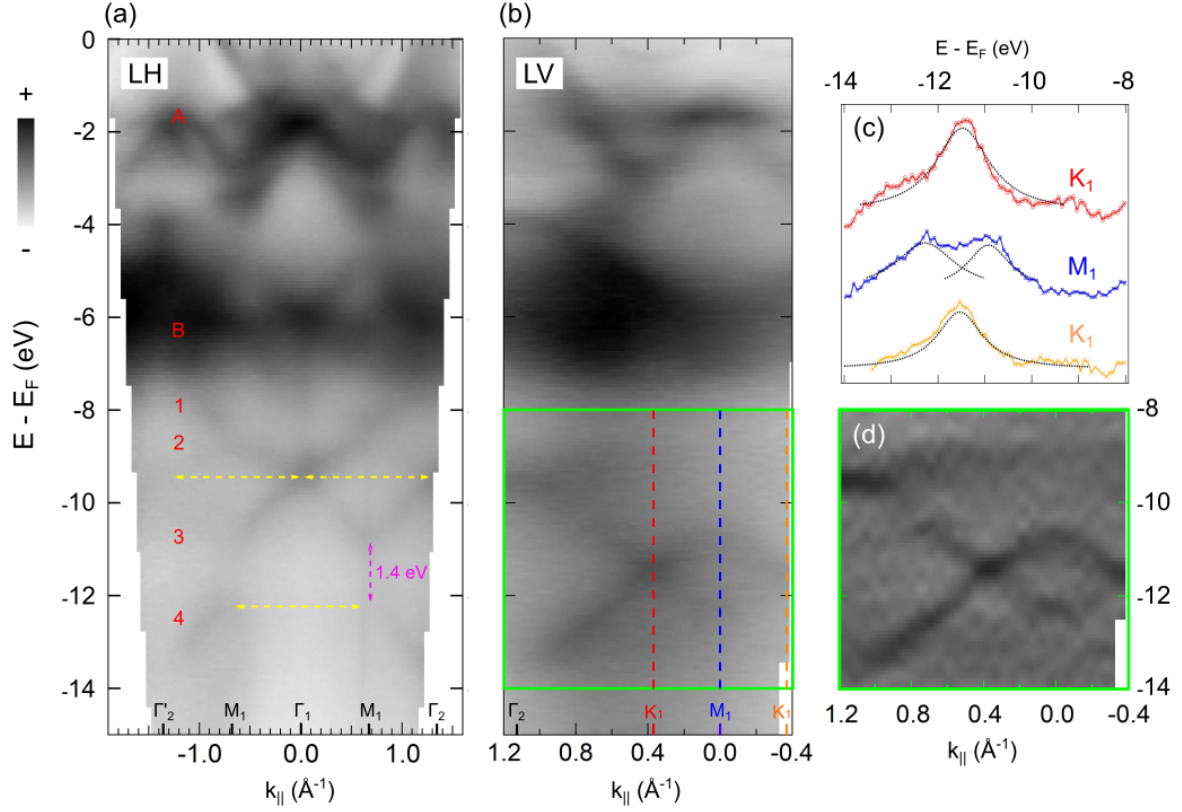

Figure S8: (a) Experimental bands dispersion for a ML silicon oxide in the $\Gamma'_2 - M_1 - \Gamma_1 - M_1 - \Gamma_2$ high symmetry direction using LH polarization and $h\nu = 40$ eV. The distances between the high symmetry points are specified in yellow. The opening of band gap at the M point is highlighted in pink. (b) Same for the $\Gamma_2 - K_1 - M_1 - K_1$ direction using in this case LV polarization. Vertical colored dashed lines correspond to EDC at high symetry points which are plotted in (c). (d) Second derivative of spectrum (b), calculated in the range of [-8,-14] eV.

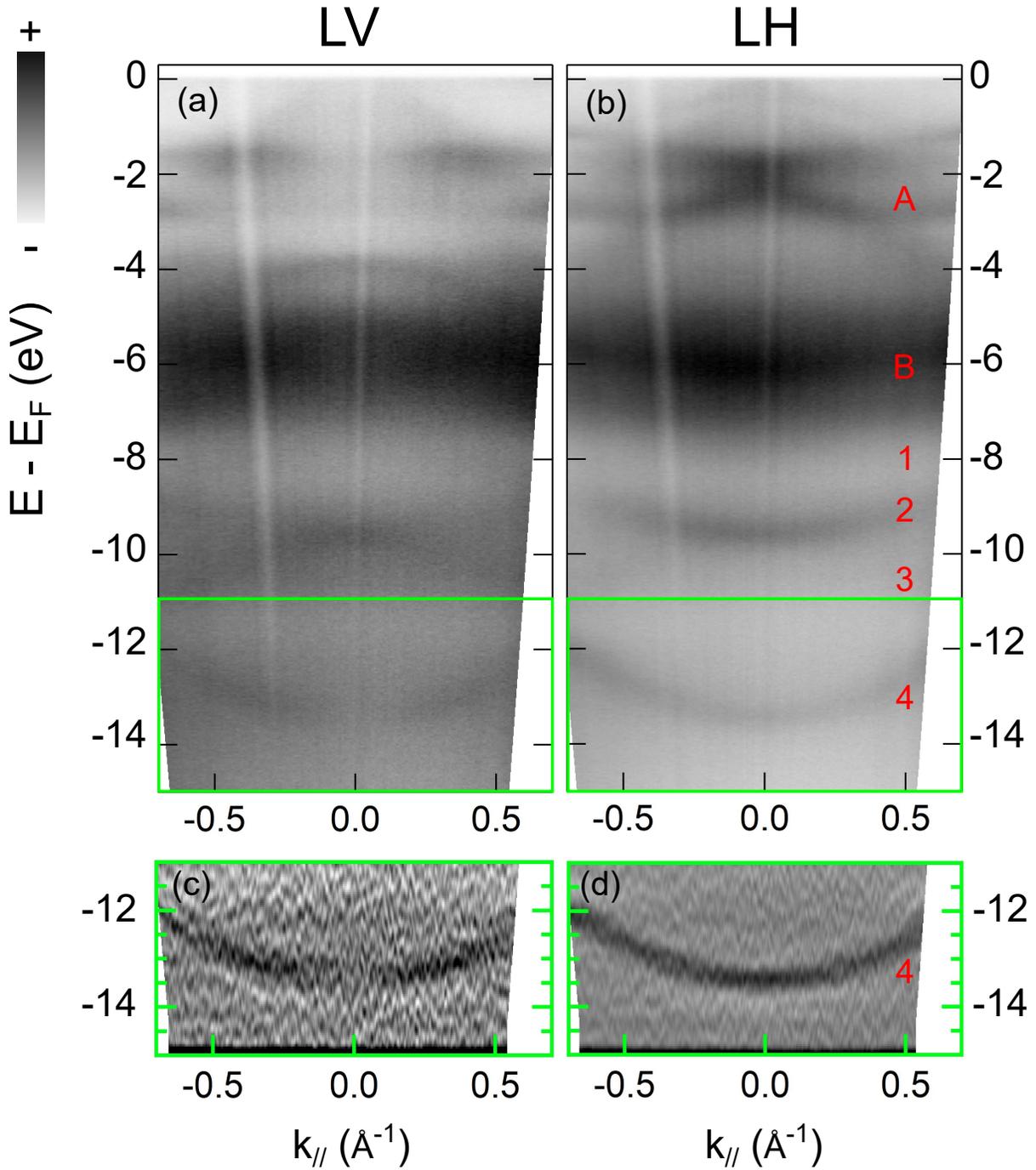

Figure S9: Angle-resolved spectra of ML silicon oxide in the second BZ for (a) LV and (b) LH polarization with $h\nu = 40$ eV. (c,d) corresponding second derivatives calculated in the energy range of [-11, -15] eV.

10

To validate our theoretical results, silicon oxide ML on a five layer slab of ruthenium was tested, using different levels of approximation for the exchange-correlation energy. We have considered local (LDA[1]) with and without the spin-orbit coupling (SOC), semilocal (PBE,[2] SCAN[3]), hybrid (HSE06[4]) and Van der Waals (optB88-vdW[5]) density functionals for the exchange correlation energy. Figure S10b shows a comparison of the electronic band dispersion along the M-Γ-K-M path calculated with the LDA (three and five layer slab) and the SCAN functional. It is worth noticing that intermediate-range Van der Waals interactions are included in the SCAN functional.

Due to the expensiveness of the hybrid HSE06 calculation, we were not able to obtain a complete band dispersion with this functional. Thus, we analyze the eigenvalue at the Γ-point corresponding to the five characteristic bands obtained with the different XC functionals (Figure S11)

In spite of the small shifts, the overall behavior does not change by considering different levels of the XC approximation. There is not gap closing between eigenvalues "3" and and "2", the latest one being three times degenerated (with eigenvalues "1" and "0" in Figure S11) at the Γ-point.

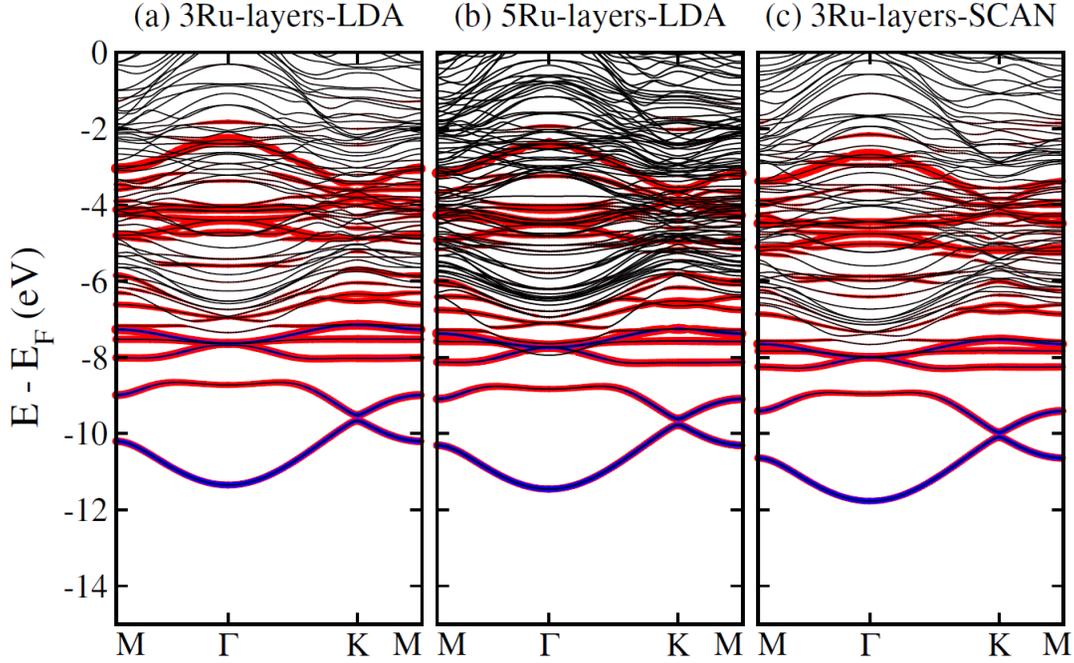

Figure S10: Comparison of electronic band dispersion for silicon oxide on top of a (a) three layer slab, (b) five layer slab calculated with the LDA functional and (c) three layer slab with the SCAN functional.

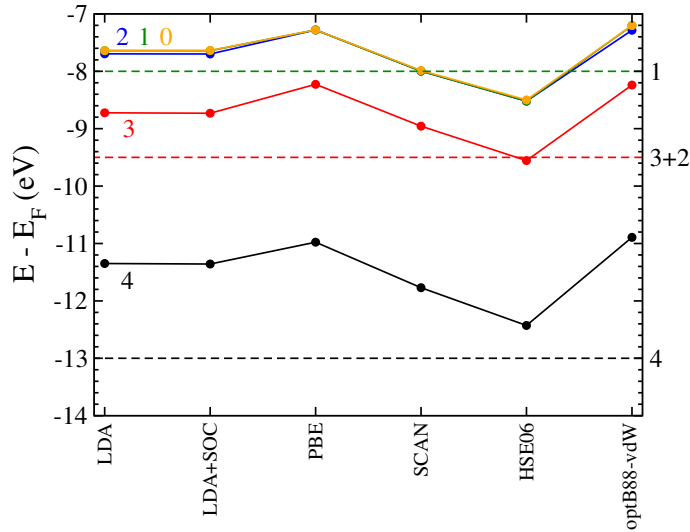

Figure S11: Eigenvalues (0 to 4) at the $\Gamma$ point for the five characteristic bands in the range [-7, -14] using local (LDA) with and without spin-orbit coupling (SOC), semilocal (PBE, SCAN), hybrid (HSE06) and Van der Waals (optB88-vdW) density functionals for the exchange correlation energy. Dashed lines (labeled at the right side) correspond to the values obtained with the ARPES measurements.

# Author Information

*E-mail: geoffroy.kremer@univ-lorraine.fr



\end{document}